\documentclass[conference,11pt]{IEEEtran}
\usepackage[dvips]{graphicx}
\usepackage{amssymb,amsmath}
\usepackage{fancyhdr}
\usepackage{lastpage}

\newcommand{\ie}{{\it i.e. }}
\newcommand{\etal}{\textit{et al. }}

\begin{document}
\title{COUNTERACTING BYZANTINE ADVERSARIES WITH NETWORK CODING:\\ AN OVERHEAD ANALYSIS}

\author{\thanks{Supported by a grant}\authorblockN{MinJi Kim}
\authorblockA{\small Massachusetts Institute of Technology\\
Cambridge, MA 02139, USA\\
Email: minjikim@mit.edu}
\and
\authorblockN{Muriel M\'{e}dard}
\authorblockA{\small Massachusetts Institute of Technology\\
Cambridge, MA 02139, USA\\
Email: medard@mit.edu}
\and
\authorblockN{Jo\~{a}o Barros}
\authorblockA{\small Instituto de Telecommunica\c{c}\~{o}es\\
Universidade do Porto, Porto, Portugal\\
Email: barros@dcc.fc.up.pt}
}
\maketitle
\begin{abstract}

\textit{
Network coding increases throughput and is robust against failures and erasures. However, since it allows mixing of information within the network, a single corrupted packet generated by a Byzantine attacker can easily contaminate the information to multiple destinations.}

%Reference \cite{fang} proposed a signature scheme for Byzantine detection on a packet-by-packet basis, and reference \cite{resilient} showed that, by augmenting polynomial hashes, Byzantine adversaries can be detected in multicast network on a generation-by-generation basis.

\textit{
In this paper, we study the transmission overhead associated with three different schemes for detecting Byzantine adversaries at a node using network coding: end-to-end error correction, packet-based Byzantine detection scheme, and generation-based Byzantine detection scheme. In end-to-end error correction, it is known that we can correct up to the min-cut between the source and destinations. However, if we use Byzantine detection schemes, we can detect polluted data, drop them, and therefore, only transmit valid data. For the dropped data, the destinations perform erasure correction, which is computationally lighter than error correction. We show that, with enough attackers present in the network, Byzantine detection schemes may improve the throughput of the network since we choose to forward only reliable information. When the probability of attack is high, a packet-based detection scheme is the most bandwidth efficient; however, when the probability of attack is low, the overhead involved with signing each packet becomes costly, and the generation-based scheme may be preferred. Finally, we characterize the tradeoff between generation size and overhead of detection in bits as the probability of attack increases in the network.
}
\end{abstract}

\thispagestyle{fancy}
\lfoot{\small{978-1-4244-2677-5/08/\$25.00 \copyright 2008 IEEE}}
\cfoot{\thepage\ of\ \pageref{LastPage}}

\section{Introduction}

Network coding, which was first introduced in \cite{ahlswede}, allows algebraic mixing of information in the intermediate nodes. This mixing has been shown to have numerous performance benefits. It is known that network coding maximizes throughput \cite{ahlswede}, as well as robustness against failures \cite{algebraic} and erasures \cite{reliable}. However, a major concern for network coding system is its vulnerability to Byzantine adversaries. A single corrupted packet generated by a Byzantine adversary can contaminate all the information to a destination, and propagate to other destinations quickly. For example, in random linear network coding \cite{reliable}, one corrupted packet in a generation can prevent a receiver from decoding any data from that generation even if all the other packets it has received are valid.

There are several papers that attempt to address this problem. One approach is to correct the errors injected by the Byzantine adversaries using \emph{network error correction} \cite{errorcorrection}. Reference \cite{errorcorrection} bounds the maximum achievable rate in an adversarial setting, and generalizes the Hamming, Gilbert-Varshamov, and Singleton bounds. Furthermore, Jaggi \etal \cite{resilient} propose a distributed, rate-optimal, network coding scheme for multicast network that is resilient in the presence of Byzantine adversaries.

However, this introduces another question: if the existence of Byzantine adversaries within the network is suspected, can we do better than just using error correction codes? Rather than just naively using error correcting codes at the destinations, can we actively detect and drop corrupted packets? If so, what kind of detection scheme should we use? In addition, with the overhead associated with Byzantine detection in coded systems, do we still outperform the non-coded solution? The goal of this paper is to answer some of these questions.

We compare the overhead associated with Byzantine detection schemes in terms of bits -- polluted packets that are sent or unpolluted packets that are dropped, as well as bits used for hashes to detect attacks -- and the amount of bandwidth saved from employing such detection schemes. The computational overhead associated with computing the hashes, or checking for corrupted packets will be dealt with elsewhere.

The paper is organized as follows. In Section \ref{sec:background}, we present the background and related material. In Section \ref{sec:networkmodel}, we introduce our network model. In Section \ref{sec:overhead}, we analyze the overhead associated with the three Byzantine detection schemes. In Section \ref{sec:tradeoffs}, we study the tradeoffs of different detection schemes, as the probability of Byzantine attack varies. Finally, we summarize our contributions of this paper.

\section{Background}\label{sec:background}
\subsection{Network Coding}\label{sec:bg-netcod}

Reference \cite{ahlswede} shows that network coding allows a source to multicast information at a rate approaching the smallest cut between the source and any receiver, as the coding symbol size approaches infinity. Li \etal \cite{LYC} show that linear network coding is sufficient for multicast networks to achieve the optimum. Subsequently, Koetter \etal \cite{algebraic} present an algebraic framework for linear network coding in arbitrary networks.

As a result, there has been a great emphasis on linear network coding. For instance, Ho \etal \cite{rlc} propose a simple, practical capacity-achieving code, in which every node construct its linear code randomly and independently from all other nodes. This simple construction has been shown to achieve capacity with probability approaching 1 exponentially fast with the field size. This result indicates that linear network coding is a simple yet a very powerful tool. Furthermore, Lun \etal \cite{reliable} show that \emph{random block linear network coding system} achieves capacity. In a random block linear network coding system, a source generates information in blocks of $G$ packets (called a \emph{generation}). The source then multicasts to its destination nodes using random linear network coding, where only the packets from the same generation are mixed. In this paper, we shall consider systems that use random block linear network coding.

It is important to note that random linear network coding is a distributed protocol, which requires no state information. It uses the encoding vector -- a vector of coefficients of the linear transformation of the original packets -- to encode/decode the original data. Therefore, random block linear network coding is resilient in dynamic/unstable networks where state information may change rapidly or may be hard to obtain.

\subsection{End-to-end error correction scheme}\label{sec:bg-errorcorrection}

In \cite{resilient}, Jaggi \etal introduce the first distributed polynomial-time rate-optimal network codes that work in the presence of Byzantine nodes and is information-theoretically secure. In their work, Jaggi \etal present algorithms that are resilient against adversaries of different capabilities. Given an adversary who can eavesdrop on all links and jam $z$ links, their algorithm achieves a rate of $C - 2z$, where $C$ is the network capacity; given an adversary who can observe only $z_e$ links and jam $z$ links where $z_e < C - 2z$, the algorithm achieves a rate of $C - z$. These rates are the maximum achievable rate given the power of the adversary.

The idea behind this algorithm is that the corrupted packets injected by the adversary can be considered as packets from a secondary source; therefore, with enough information at the destinations, the destination nodes can decode both the legitimate source's packets as well as the adversary's packets. To allow the destination nodes to distinguish the legitimate packets from the adversary's packets, the source judiciously adds redundancy such that the adversary's packets cannot satisfy the constraints imposed by the redundancy. For instance, the source may add redundancy so that a certain functions evaluate to zero on the source's packets.

\subsection{Packet-based Byzantine detection scheme}\label{sec:bg-packet}

There are several signature schemes that have been presented in the literature. For instance, \cite{homomorphic1}\cite{homomorephic2} use homomorphic hash functions to detect polluted packets. In addition, Charles \etal \cite{elliptic} propose a signature scheme for network coding based on Weil pairing on elliptic curves. Although this scheme does not require a secure channel, it is computationally expensive. In \cite{fang}, Fang \etal propose a signature scheme for network coding, which makes use of the linearity property of the packets in a coded system. This scheme does not require intermediate nodes to decode coded packets to check the validity of a packet; therefore, it is efficient in terms of computational cost as well as delay. In this paper, we shall consider the signature scheme from \cite{fang}, which is designed for transmitting large files that are broken into blocks viewed as vectors. Taking advantage of the fact that in linear network coding, any valid packet transmitted should belong to the subspace spanned by the original set of vectors, Fang \etal design a signature that can be used to easily check the membership of a received vector in the given subspace, while making it hard to generate a fake signature that is not in the subspace but passes the check.

The overhead associated with the signature scheme is also analyzed in \cite{fang}. Given a file, we break it into $m$ blocks/vectors, which are elements in $n$-dimensional vector space $\mathbb{F}^n_p$. Then, the size of the file, a vector, and the coding vectors are $mn\log{p}$, $n\log{p}$, and $(m+n)\log{p}$, respectively. This scheme requires a public key $\mathbf{K}$ of size at least $(m+n)\log{q}$, where $q$ is a large prime number such that $p$ is a divisor of $(q-1)$. In a typical cryptographic applications, $p$ is 160 bits, $q$ 1024 bits, and $\mathbf{K}$ approximately $6(m+n)/mn$ times the file size. It is important to note that although the overhead of the signature is quite small (less than 0.1\% of the file size), the overhead of the key distribution can be quite large. The cost of key distribution is not considered in \cite{fang}.

This detection scheme assumes no knowledge of the bandwidth available between the source and the destinations -- therefore, can validate packets under varying degree of attack. Thus, this algorithm achieves rate of $C - z$ minus the overhead of the signature scheme, where $C$ is the network capacity, and $z$ is the rate at which the adversary can inject packets into the network.

%\vspace*{0.5cm}
%\begin{figure*}[tbp]
%\centering
%\includegraphics[width=0.85\textwidth]{trustednode} \caption{Diagram of the network and a non-malicious node $v$}\label{fig:trustednode}
%\end{figure*}

\subsection{Generation-based Byzantine detection scheme}\label{sec:bg-generation}

\begin{figure}[tbp]
\centering
\hspace*{-.7cm}
\includegraphics[width=0.55\textwidth]{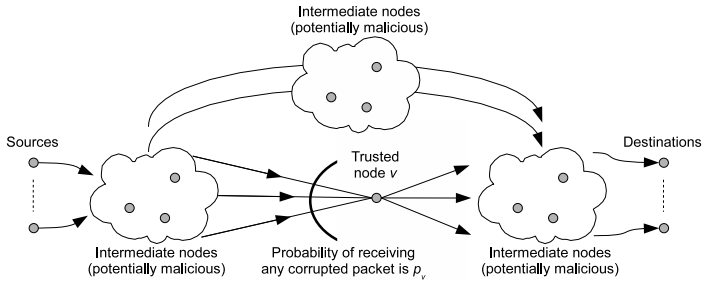} \caption{Diagram of the network and a non-malicious node $v$}\label{fig:trustednode}
\vspace*{-.5cm}
\end{figure}

In \cite{detection}, Ho \etal introduce an information-theoretic approach for detecting Byzantine adversaries, which only assumes that the adversary did not see all linear combinations of the source packets received by the destinations. They propose a scheme whose detection probability varies with the length of the hash $k$, coding field size $q$, and the amount of information about the random code unknown to the adversary $s$. A polynomial hash of a flexible length is augmented to each packet in the generation. Once the destination node receives enough packets to decode a generation, the destination node can decode the set and detect error with probability at most $\left(\frac{k+1}{q}\right)^s$. The intuition behind this scheme is that if a packet is reliable, then its data and hash symbols are consistent with its coefficient vector; and a linear combination of reliable packets is also reliable. In \cite{detection}, Ho \etal design a hash function, which is both efficient to compute and difficult to forge without seeing all valid packets and their coding coefficients.

This generation based scheme is very cheap and sensitive. For example, with 2\% overhead ($k=50$), $\log{q} = 7$, $s = 5$, the detection probability is at least 98.9\%; with 1\% overhead ($k=100$), $\log{q} = 8$, $s=5$, the detection probability is at least 99.0\%. Furthermore, this scheme does not require any key agreement/distribution; thus, making it much cheaper than the packet-based scheme. However, this is a block code, unlike the signature scheme from Section \ref{sec:bg-packet}; therefore, will require a priori decision on the rate. In addition, the detection can only occur at a node with enough packets from a generation -- thus, can incur large delays.

\section{Network Model}\label{sec:networkmodel}

We model the network by a directed graph $G = (V, E)$, where $V$ is the set of nodes in the network, and $E$ is the set of communication links. There are subsets of nodes, $S$ and $T \subseteq V$, which wish to send and receive data respectively.

%There exists a set $R \subseteq V$ of \emph{trusted nodes}. Trusted nodes of a network are given access to the public key of the Byzantine detection scheme in use. For example, for the packet-based Byzantine detection scheme from Section \ref{sec:bg-packet}, the trusted nodes are given the public key $\mathbf{K}$; for the generation-based Byzantine detection scheme from Section \ref{sec:bg-generation}, the trusted nodes are allowed to decode a generation, if all the packets of the given generation goes through the trusted node. Therefore, a trusted node can verify the validity of the packet or the generation. By definition, $S \subseteq R$ and $T \subseteq R$.

In this paper, we consider a non-malicious node $v\in R$. Node $v$ wishes to check the validity of the packet/generation that it forwards. For the packet-based Byzantine detection scheme from Section \ref{sec:bg-packet}, $v$ is given the public key $\mathbf{K}$; for the generation-based Byzantine detection scheme from Section \ref{sec:bg-generation}, $v$ is allowed to decode a generation, if all the packets of the given generation goes through it.

Assume that node $v$ receives $m$ packets ($n$ bits each) per time unit. If it detects an error/attack, then $v$ discards that data; otherwise, the $v$ acts like any other node in the network and forwards the data.

A key parameter to consider in this work is the probability $p_v$ of an attack to the node $v$. At any given time, the probability that $v$ receives a packet modified by the Byzantine attacker is $p_v$ as shown in Figure \ref{fig:trustednode}. Therefore, the expected number of corrupted packets that $v$ receives is $mp_v$. It is important to note that the probability $p_v$ of an attack is topology dependent (in terms of the location of $v$, Byzantine attackers, as well as the source nodes); thus, $p_v$ may be different for each node $v$. We shall assume that there is an external model of vulnerability which will give an estimate of $p_v$.

In the remaining of the paper, we shall focus on a single non-malicious node $v$, and the overhead associated with the detection schemes through this one node. This is a reasonable approach, since we are not concerned with how a malicious node uses its bandwidth. Since we are considering a specific node $v$ from now on, we will denote $p_v$ as $p$, unless specified otherwise. In the next sections, we compare the cost and the benefit of employing Byzantine detection schemes with varying $p$.

\section{Overhead analysis}\label{sec:overhead}

\subsection{Overhead of end-to-end error correction}\label{sec:nodetection}

In an end-to-end error correction scheme, the network can achieve rate of $C-z$ where $C$ is the network capacity and the adversary can jam $z$ links. Therefore, end-to-end error correction can correct up to the \emph{reliable} min-cut of the source-destinations. Thus, using this scheme, as long as the attack is within the network capacity, the intermediate nodes can transmit at the remaining network capacity; and the destination nodes employ error correction to retrieve reliable packets. It is important to note that error correction is computationally more expensive than erasure correction, which shall be used when Byzantine detection schemes are used.

In this scheme, node $v$ acts as a normal intermediate node -- it does not check the validity of the packet it has received. It just naively performs random linear network coding and forwards the information it has received. Therefore, in terms of bits transmitted at $v$, we lose on average $mnp$-bits, where $n$ is the length of a packet, $m$ is number of packets $v$ receives per time unit, and $p$ is the probability that any packet $v$ receives is corrupted. Therefore, the ratio between the overhead or corrupted bits transmitted and total bits received is:
\begin{equation}\label{eq:errorcorrection}
\frac{mnp}{mn} = p.
\end{equation}

\subsection{Overhead of Byzantine detection for packets}\label{sec:packets}

In this section, we analyze the overhead associated with the packet-based Byzantine detection algorithm (Section \ref{sec:bg-packet}). We consider the scenario where node $v$ checks the validity of every packet it receives using the public key $\mathbf{K}$ available to it. Then, $v$ uses random linear network coding to forward only the packets that passed the check. If $v$ detects an error in a packet, then it discards it -- by doing so, $v$ does not waste its bandwidth in transmitting corrupted data with high probability, and destination nodes perform erasure correction on the packets that have been dropped, which is computationally cheaper than error correction required for the error correction scheme in Section \ref{sec:nodetection}. However, each packet needs to contain enough information (a polynomial hash) that can be used to verify its integrity. This overhead associated with the Byzantine detection scheme reduces the rate at which data is transmitted.

Assume that this polynomial hash is of size $h_p$ bits per packet. Before analyzing the cost/benefit of the packet-based detection scheme, we first study the rate of actual data transmission. Since $p$-fraction of packets received by $v$ is erroneous, $v$ only forwards in total $mn - mnp$ bits per unit time, of which only $(1-\frac{h_p}{n})$ fraction of the bits are actual data bits. Thus, the ratio between the actual data bits transmitted and total bits transmitted is:
\[
\frac{(mn-mnp)(1-\frac{h_p}{n})}{mn-mnp} = 1 - \frac{h_p}{n}.
\]
This result is not a surprising one -- as it shows that when we employ a Byzantine detection scheme, $v$ can filter out all the corrupted packets. As a result, every bit $v$ transmits is a valid data bit. This analysis also extends to that of generation-based Byzantine detection scheme in Section \ref{sec:generations}.

The overhead associated with the packet-based Byzantine detection scheme can be analyzed as follows. By discarding the corrupted packets, node $v$ can on average save its bandwidth by $mnp$ bits per unit time at a cost of $h_p$ bits per packet. Therefore, the expected cost of the this scheme at $v$ is:  $\max\{0, m(h_p - np)\}$ bits per time unit. Therefore, the ratio between the overhead and the total bits received is:
\begin{equation}\label{eq:packet}
\frac{\max\{0, m(h_p - np)\}}{mn} = \frac{\max\{0, h_p - np\}}{n}.
\end{equation}

When the probability of error $p$ is high enough,  then checking each packet for error saves on bits transmitted -- \ie $m(h_p - np) < 0$, which shows that the cost of the signature scheme is canceled by the bandwidth gained from dropping the corrupted packets, and thus, reduces the overall cost to zero. Therefore, this approach is the most sensible when the network is expected to be unreliable or under heavy attack.

It is important to note that the packet-based Byzantine detection scheme assumes the presence of a public key distribution infrastructure, the details of which we are not concerned with here. However, there are transmission as well as computational overhead associated with such an assumption, which has been studied in \cite{secretKey}\cite{publickeyTinyOS}. Furthermore, this packet-based detection scheme is designed for a large file which is broken into small blocks. Therefore, every new file requires a new public key; the cost of key distribution becomes a significant part of the overhead of this signature scheme. Thus, depending on the public key distribution infrastructure used and the frequency of key renewal, the packet-based detection scheme will incur a much higher overhead. This would result in shifting the transmission overhead in Figure \ref{fig:comparison} outwards.

\subsection{Overhead of Byzantine detection for generations}\label{sec:generations}

In this section, we shall assume our system uses random block linear network coding with generation size $G$. In this generation-based Byzantine detection scheme, a node checks for possible error/attack on a generation after collecting enough packets from the generation. If the node detects an error, then it discards the entire generation of $G$ packets; otherwise, it forwards the data. The destination nodes perform erasure correction on the generations that have been dropped, which is computationally cheaper than error correction required in Section \ref{sec:nodetection}.

Thus, this scheme requires only one hash for the entire generation --- saving bits on the hashes compared to the packet-based detection scheme. However, one corrupted packet in a generation can make a node drop the entire generation and make the network inefficient.

For a more detailed analysis on the overhead associated with this scheme, assume that the hash is of size $h_g$ bits per generation. The probability of dropping a generation of $G$ packets is given by:
\begin{align*}
p_g&=\Pr(\text{generation dropped})\\ & = 1 - \Pr(\text{All $G$ packets are valid})\\
&= 1- (1-p)^G.
\end{align*}

Therefore, the probability that a generation is forwarded by $v$ is $1 - p_g = (1-p)^G$; node $v$ is expected to transmit $(1-p)^G nG$ bits per unit time. By similar analysis as in Section \ref{sec:packets}, the fraction of actual data bits of the $(1-p)^G nG$ transmitted bits is $1 - \frac{h_g}{nG}$.

The overhead associated with this scheme includes the hash of $h_g$ bits per generation. However, unlike the packet-based detection scheme, this scheme may drop uncorrupted packets along with the corrupted packets, if a generation is deemed corrupted. Therefore, the overhead of the generation-based detection scheme also needs to include the uncorrupted packets that were dropped. The expected number of uncorrupted bits in a generation is given by: $(1-p)nG$ bits. Therefore, the expected uncorrupted bits dropped per generation is: $p_g(1-p)nG$ bits.

However, this scheme saves on bandwidth by dropping generations with corrupted packets. On average, there are $pnG$ corrupted bits in a each generation. Therefore, the expected overall overhead (in bits) of Byzantine detection per generation is:
\[
\max\{ 0, h_g + p_g(1-p)nG - pnG\} \text{ bits.}
\]
Thus, the ratio between the overhead and the total bits received is:
\begin{equation}\label{eq:generation}
\frac{\max\{0, h_g + p_g(1-p)nG - pnG\}}{nG}.
\end{equation}

It is important to note that, for this scheme to work, $v$ needs to receive at least $G$ packets from each generation so that it can decode the generation and use the generation-based Byzantine detection scheme to detect attackers. This may seem to indicate that this scheme is only applicable as an end-to-end Byzantine detection scheme or requires that $v$ receives all packets, but it can be used as a \emph{local} Byzantine detection scheme.

\begin{figure}[h!]
\begin{center}\hspace*{-.5cm}
\includegraphics[width=0.50\textwidth]{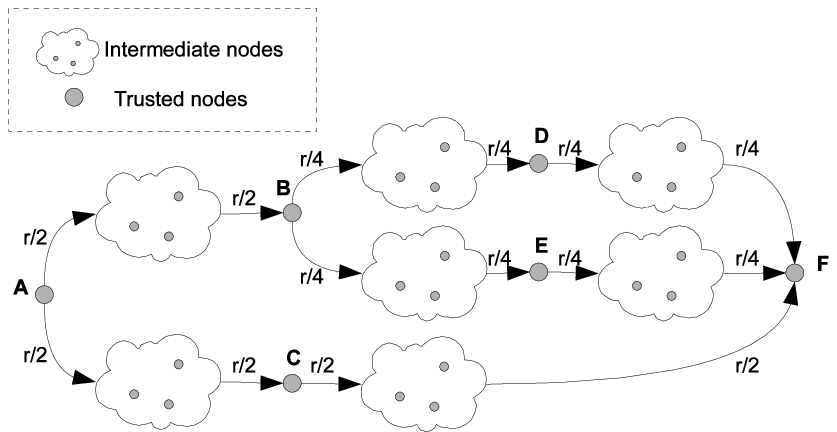}\end{center} \vspace*{-.5cm}\caption{Network with non-malicious nodes $\mathbf{A}$, $\mathbf{B}$, $\mathbf{C}$, $\mathbf{D}$, $\mathbf{E}$, and $\mathbf{F}$ where node $\mathbf{A}$ is transmitting at a total rate of $\mathbf{r}$ to node $\mathbf{F}$}
\label{fig:generationsplit}
\vspace*{-.5cm}
\end{figure}

Consider a network with non-malicious nodes $\mathbf{A}$, $\mathbf{B}$, $\mathbf{C}$, $\mathbf{D}$, $\mathbf{E}$, and $\mathbf{F}$, as shown in Figure \ref{fig:generationsplit}, and assume that this network uses the generation-based Byzantine detection scheme with generation size $G$. In this network, node $\mathbf{A}$ is transmitting data at rate $\mathbf{r}$ to node $\mathbf{F}$; however, $\mathbf{A}$ sends half of its data through $\mathbf{B}$ and the other half through $\mathbf{C}$. It may seem that nodes $\mathbf{B}$ and $\mathbf{C}$ cannot check the validity of any generation transmitted by $\mathbf{A}$ since it is unlikely that they will receive enough packets from any generation; however, $\mathbf{B}$ and $\mathbf{C}$ can check the validity of the \emph{sub-generation} they receive, where by sub-generation, we mean a collection of $G/2$ encoded packets from $\mathbf{A}$. By a similar argument, $\mathbf{D}$, $\mathbf{E}$, and $\mathbf{F}$ can check the validity of a sub-generation of $G/4$, $G/4$, and $G$ packets from $\mathbf{A}$, respectively. Therefore, a node can check every sub-generation it forwards.

\section{Trade-offs}\label{sec:tradeoffs}

In this section, we choose $h_p = \frac{6}{100}n$ and $h_g = \frac{2}{100}nG$. As noted in Section \ref{sec:bg-packet}, the signature and the public key used in the packet-based scheme are approximately 0.1\% and $6(m+n)/mn\approx 6\%$ of the data file size. In any public key infrastructure, the public key has to be transmitted at least once, therefore, we make an underestimate of the cost and choose $h_p = \frac{6}{100}n$. From Section \ref{sec:bg-generation}, the hash for the generation-based scheme is approximately 2\% of the transmitted bits.

\subsection{A comparison of coded and non-coded systems}\label{sec:nocoding}

A non-coded system, unlike its coded counterpart, requires state information such as network topology or buffer information. As a result, the most effective attack on a routing network is an attack on the control traffic. However, as we noted in Section \ref{sec:bg-netcod}, network coding systems are robust against such an attack; making it especially more robust in dynamic/unstable networks.

In addition, for an effective Byzantine detection in a routing network, we need all nodes in the network to be authenticated; therefore, each packet would need a signature as well as a hash to verify the identity of the sender and the content of the packet. Therefore, a non-coded system would incur overhead similar to that of packet-based scheme without the benefit of throughput gain due to network coding. There are various literature on the overhead analysis of secure routing protocols, especially for wireless ad hoc networks \cite{hubaux}\cite{marti}. In \cite{marti}, it has been shown that these routing protocols can incur up to 24\% overhead; making the cost of detection non-negligible without the performance benefits of coding.

\subsection{Generation size $G$ in the generation-based scheme}

\begin{figure}[h!]
\begin{center}\hspace*{-.5cm}
\includegraphics[width=0.55\textwidth]{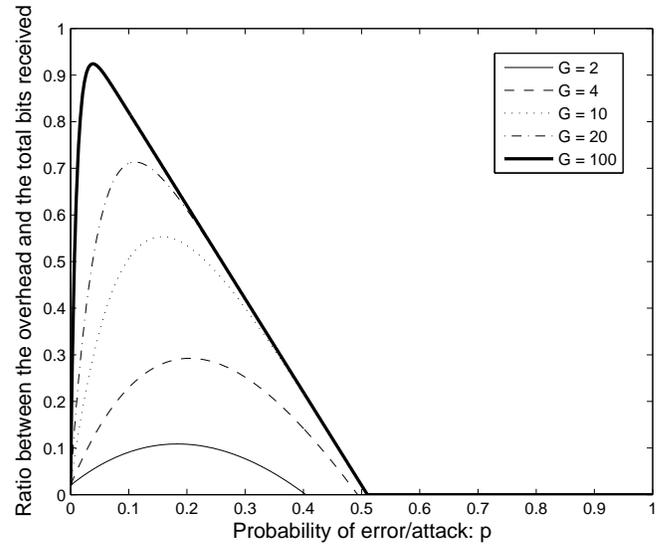}\end{center}\vspace*{-.5cm} \caption{Ratio between the expected overhead and the total bits received by a node for generation-based detection with generation size $G$, packet size $n = 1000$, and hash size $h_g = \frac{2}{100}nG$}\label{fig:generations}
\vspace*{-.5cm}
\end{figure}

In Figure \ref{fig:generations}, we see that given an error probability $p$, as generation size increases, the cost of the generation-based scheme increases dramatically. If the generation size $G$ is large enough, there will be at least one corrupted packet in a generation with high probability even for small $p$. This can be easily verified with an asymptotic analysis of Equation \ref{eq:generation} as $G \rightarrow \infty$:
\begin{align*}
\lim_{G\rightarrow \infty} &\frac{\max\{0, h_g + p_g(1-p)nG - pnG\}}{nG}\\
& \rightarrow \max\{0,1-2p\}.
\end{align*}

This indicates that a large generation size is undesirable -- as almost every generation is found faulty and dropped; making the network throughput to zero. However, this constraint should not become a relevant limiting factor in many MANET systems, since the generation sizes are kept small to keep the coding/decoding cost low.

Another interesting thing to note in Figure \ref{fig:generations} is that the cost peaks at a probability $p \approx 0.2$. At $p \approx 0.2$, the generation-based scheme drops many generations for a few corrupted packets each generation; as a result, it drops many valid packets to filter out a few corrupted packets. Thus, at a moderate rate of attack, the generation-based scheme suffers. When $p << 0.2$, the generation-based scheme does well -- since the probability of error is low and we distribute the cost of hashing across an entire generation, we do not waste bandwidth. When $0.2 << p < 0.5$, this scheme blocks all generation that is corrupted -- and as a result, the scheme does not waste bandwidth on corrupted packets. Nodes transmit only valid information; however, this happens rarely. This also means that we do not use the bandwidth to transmit ``good'' packets since we throw them away along with the corrupted packets in the generation. When $p > 0.5$, the expected throughput is near zero; therefore, the expected cost is zero since we do not transmit any corrupted (as well as good) bits.

\subsection{A comparison of the three schemes}

%\begin{figure*}[btp]
%\centering
%\includegraphics[width=1.0\textwidth]{comparison2} \caption{Expected cost in bits transmitted for the three schemes}\label{fig:comparison}
%\end{figure*}

\begin{figure}[tbp]
\vspace*{-.7cm}
\begin{center}\hspace*{-.5cm}
\includegraphics[width=0.53\textwidth]{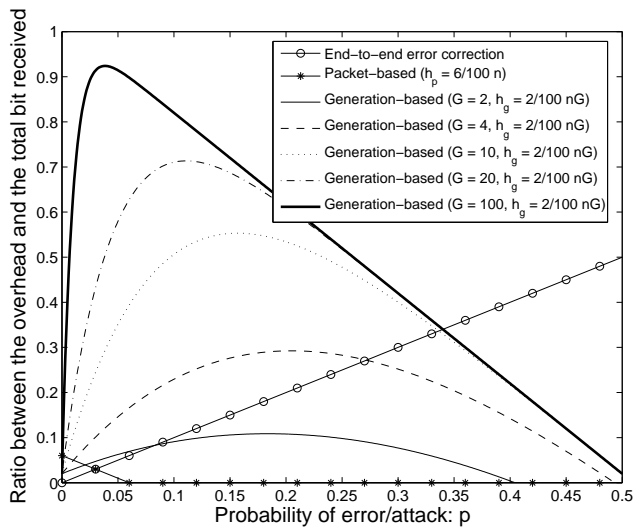}\end{center}\vspace*{-.5cm} \caption{Ratio between the expected overhead and the total bits received by a node with $h_p = \frac{6}{100}n$, $h_g= \frac{2}{100}nG$}
\label{fig:comparison}
\vspace*{-.6cm}
\end{figure}

\begin{figure}[tbp]
\vspace*{-.7cm}
\begin{center}\hspace*{-.5cm}
\includegraphics[width=0.53\textwidth]{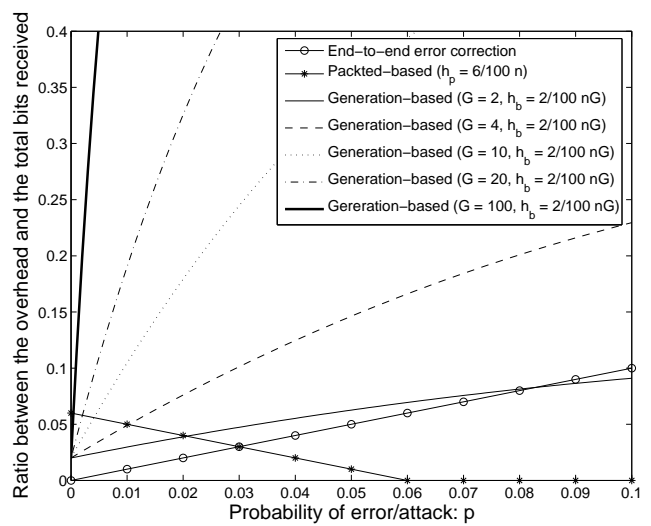}\end{center}\vspace*{-.5cm} \caption{Ratio between the expected overhead and the total bits received by a node with $h_p = \frac{6}{100}n$, $h_g= \frac{2}{100}nG$ for $p \in [0, 0.1]$}
\label{fig:blownup}
\vspace*{-.6cm}
\end{figure}

In Figures \ref{fig:comparison} and \ref{fig:blownup}, we compare the three schemes. As mentioned in Section \ref{sec:nodetection}, the expected cost of error correction scheme is linearly proportional to the error probability. Therefore, when $p$ is large, this scheme performs badly. However, this simple scheme where a node ignores any error/attacks and just forwards all the information it receives outperforms the detection schemes when $p$ is low ($p <0.03$). When the probability of error is small, the cost of detection is an overhead that exceeds the cost introduced by the attackers.

It is interesting to compare the packet-based and the generation-based scheme. When the probability of error is low, the overhead of hash is costly for the packet-based scheme, since it is devoting $h_p$ bits per packet to detect an unlikely attack. In such a setting, the generation-based scheme performs well --- it distributes the cost of the hash ($h_g$ bits) over $G$ packets and still detects the few attacks that it may encounter. However, as the probability of attack increases, the cost of hashes become ``cheaper'' since the bandwidth wasted by transmitting a corrupted packets increases. This is where the packet-based scheme outperforms the generation-based scheme. However, it is important to note that we underestimate the overhead associated with the packet-based scheme in this paper as we do not take into account the public key distribution cost, which the generation-based scheme does not require.

\section{Conclusions}\label{sec:summary}

When there are enough attackers present in the network, Byzantine detection (either packet-based or generation-based) improves the throughput since we can choose to only forward information that is clean -- rather than send the corrupted information and have the destination correct it.

As mentioned in Section \ref{sec:tradeoffs}, if the probability of receiving a bad packet is high enough (i.e. the number of corrupted bits is higher than the cost of checking each packet), then the packet-based Byzantine detection is the most bandwidth efficient. However, if the probability of an attack is low, then the overhead of attaching a hash for every packet becomes costly. Therefore, the generation approach in Section \ref{sec:generations} is appropriate. However, although the generations allow us to reduce the cost of the hash, the probability that a corrupted packet is in a generation increases with the size of the generation. Therefore, a right balance between the generation size and the error probability is needed if we choose to use generation-based Byzantine detection.

\section*{Acknowledgement}
This material is based upon work under a subcontract \#069145 issued by BAE Systems National Security Solutions, Inc. and supported by the DARPA and the Space and Naval Warfare System Center, San Diego under Contract No. N66001-08-C-2013.

\bibliographystyle{IEEEtran}
\bibliography{References}

\end{document}